\documentstyle{ichep}
\def \DZ {D\O }
\def\etmiss {\mbox{${\hbox{$E$\kern-0.6em\lower-.1ex\hbox{/}}}_t$}}
\def\Zone{\ifmmode{\tilde{Z_1}}\else{$\tilde{Z_1}$}\fi}
\begin{document}
\title{Search for Supersymmetry and \\
Leptoquark States at FNAL}

\author{Sharon Hagopian}

\affil{The Florida State University\\
Department of Physics\\
Tallahassee, Florida U.S.A}

\abstract{ Searches have been made
for first generation scalar and vector leptoquarks by the \DZ\space
collaboration and for second generation scalar leptoquarks by the
CDF collaboration.
The data sample is from the 1992-93 $p\bar p$ run at
$\sqrt s=1.8$ TeV  at the Fermilab Tevatron Collider.
Assuming that leptoquarks are pair produced and
decay into charged leptons and quarks with branching fraction $\beta$,
mass limits at the 95\% Confidence Level (CL) have been
obtained. For first generation scalar leptoquarks
the lower mass limit is 130 GeV/c$^2$ for $\beta=1.0$
and 116 GeV/c$^2$ for $\beta=0.5$.
For first generation vector leptoquarks with $\kappa$, the
anomalous coupling, of 1.0 and $\beta=1.0$, the lower mass limit
is 240 GeV/c$^2$ and for $\kappa=1.0, \beta=0.5$, the lower mass limit
is 240 GeV/c$^2$.
For $\kappa=0$ and $\beta=1.0$, the lower mass limits is  190 GeV/c$^2$ and
for $\kappa=0$,
$\beta=0.5$, the lower mass limit is 185 GeV/c$^2$.
For second generation scalar leptoquarks, the mass limits are
133 GeV/c$^2$ for $\beta=1.0$  and 98 GeV/c$^2$ for $\beta=0.5$.
A search for squarks and gluinos, predicted by Supersymmetric
models, was made by \DZ\space in the three or more jets plus
\etmiss\space channel. The number of events observed was
consistent with background. For heavy squarks, a lower gluino mass limit
of 146 GeV/c$^2$ was obtained, and for equal squark and gluino masses
a mass limit of 205 GeV/c$^2$ was obtained at the 95\% CL.}
\twocolumn[\maketitle]

\section{Introduction}
    The discovery of new particles not contained in the Standard Model (SM)
would help in choosing among the many extensions to the Standard Model
have been proposed. This report will discuss searches for two types of new
particles: leptoquarks and supersymmetric particles using data from
the 1992-1993 Fermilab $p\bar p$ collider run at $\sqrt s=1.8$ TeV.

\section{Leptoquarks}
    Leptoquarks (LQ) are exotic particles which have both lepton and
baryon quantum numbers [1]. They appear in extended gauge theories and
composite models, and can be scalar or vector particles depending on
the model [2]. Leptoquarks
link quark and lepton multiplets of the same
generation. They would be easily pair-produced at $p\bar p$ colliders, with
a production cross section that depends only slightly on the unknown coupling
$\lambda$ of the  leptoquark to ordinary leptons and quarks [3]. Leptoquarks
decay into a charged lepton and a quark, with branching
fraction $\beta$, or  into a neutrino  and a quark, with branching fraction
$(1-\beta)$.

\section{The \DZ\space Detector}

    The \DZ\space detector is described in detail elsewhere [4]. It has
uranium-liquid argon calorimeters which provide very nearly hermetic coverage
for good  \etmiss \ measurement and good hadronic and electromagnetic
resolution for good electron and jet energy measurement. It also has a central
tracking system and a muon spectrometer with coverage at large and small
angles.

\section{First Generation Leptoquark Search in \DZ}
First generation leptoquarks would decay into an electron and a quark
or into an electron neutrino
and a quark. Two possible experimental signatures for their pair
production would be:
\begin{itemize}
           \item [1)] two electrons + two jets
           \item [2)] one electron + \etmiss \ + two jets
\end{itemize}
    The \DZ\space sample for the first channel consisted of 14,780 events
with two electromagnetic clusters with $E_t>15$ GeV, from an integrated
luminosity of $13.4\pm1.6 \ {\rm pb}^{-1}$.

The offline requirements were:
\begin{itemize}
        \item [1)] two electrons with $E_t>25$ GeV passing good electron
              quality cuts
        \item [2)] two jets with $E_t>25$ GeV passing jet quality cuts
\end{itemize}
    All nine events passing these requirements have $M(ee)$ near the $Z$
 mass.  No events remain after making a 10 GeV/c$^2$ cut around the $Z$ mass.
The main background for this channel is Drell-Yan production of two electrons,
mainly at the $Z$ resonance, with two jets.
The estimated background of Drell-Yan + two jet events with
$M(ee)$ outside this mass region is 0.3 events.

    The sample for the second channel consisted of 11,480 events.
The offline requirements were:
\begin{itemize}
       \item [1)] one electron with $E_t>20$ GeV passing jet
              quality cuts
       \item [2)] two jets with $E_t>20$ GeV passing good electron
              quality cuts
       \item [3)] \etmiss $>40$ GeV
       \item [4)] transverse mass $(e,\etmiss) > 105$ GeV/c$^2$
       \item [5)] no jet-\etmiss \ correlations
\end{itemize}
No events remained after these cuts. The estimated background from
$W +$ two jet and QCD events was 0.9 events.

\begin{figure}[htp]
\vspace*{ 13pc}
 \caption{\label{pic1}\DZ\space 95\% confidence level lower limit on the first
generation  scalar leptoquark mass as a function of $\beta$.}
\end{figure}
\includegraphics{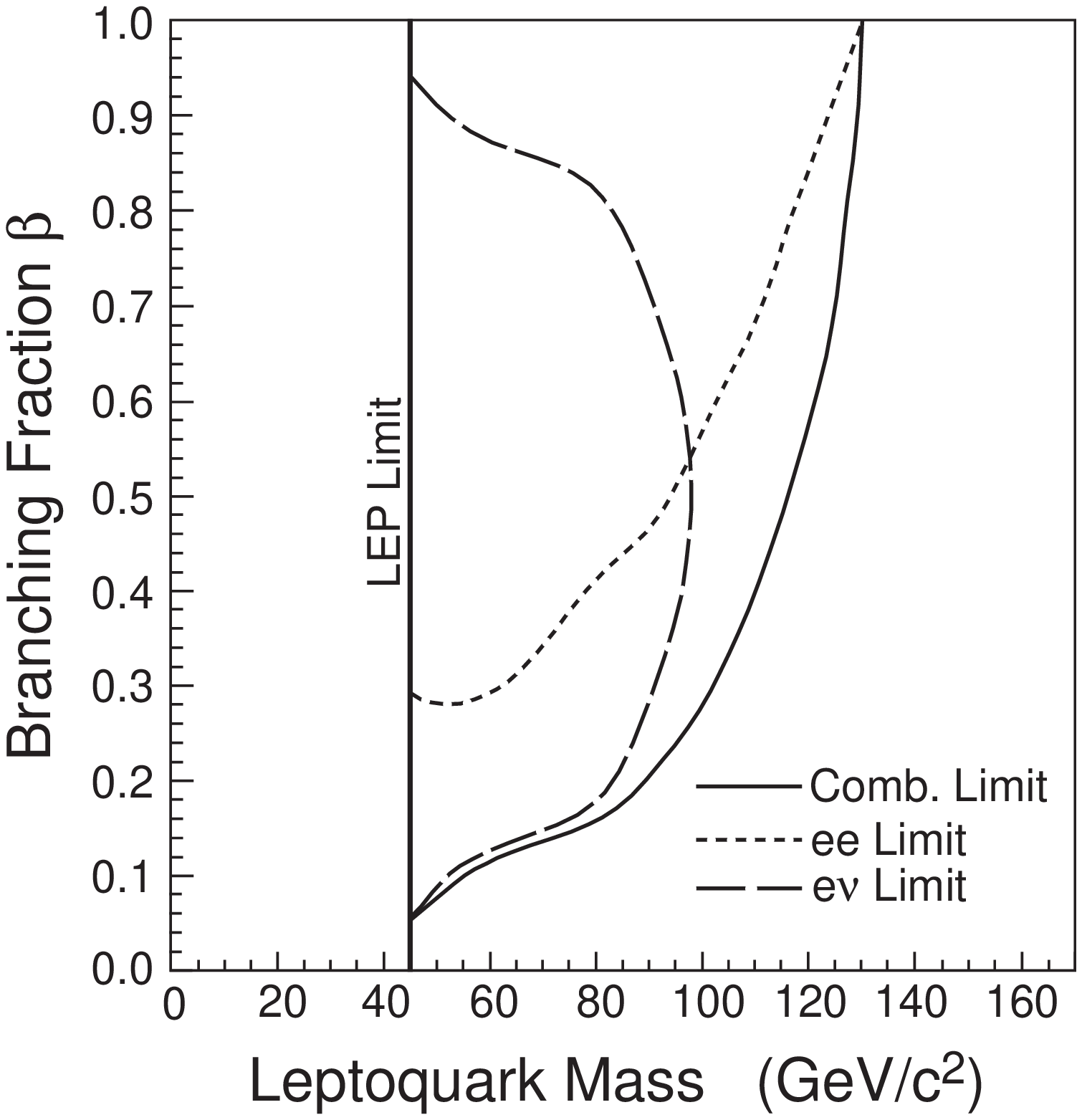}    %

\section{First Generation Scalar Leptoquark Mass Limits}

A limit on
the leptoquark mass as a function of the branching fraction $\beta$ can be
calculated by comparing the experimental cross section limit with
the theoretical prediction. The 95\% CL upper limit of the experimental
cross section can be obtained
from the observation of zero events in each channel, the luminosity, and the
\DZ\space detection efficiency for these two channels, which is a function of
leptoquark mass. For a mass of 130 GeV/c$^2$, the efficiency for the 2 electron
channel was 12\% and for the 1 electron + \etmiss\space channel it was 9\%.
The 95\% CL upper limit experimental
cross section was compared with the theoretically predicted
cross section obtained using ISAJET [5] with the
MT-LO parton distribution functions [6] to determine the lower mass limit
as a function of $\beta$.  Figure 1
shows the \DZ\space 95\% confidence limit on the mass of the first generation
scalar leptoquark. The LQ lower mass limit for
$\beta=1$ is 130 GeV/c$^2$ and for $\beta=0.5$ is 116 GeV/c$^2$. These limits
differ from those previously published by \DZ\space[7]
due to a revision of the luminosity calculation.

\section{First Generation Vector Leptoquark Mass Limits}
Recently, a cross section calculation for
the pair production of vector leptoquarks has become available
[8]. Since vector leptoquarks are composite particles
in some models, their production cross section depends on $\kappa$, the
anomalous coupling parameter, where $\kappa=1$
for gauge coupling and $\kappa=0$ for maximum
anomalous coupling. Assuming the same detection efficiency for
vector leptoquarks as for scalar leptoquarks, and using the calculated
vector LQ cross section, one can translate the results of the scalar
LQ search into mass limits for vector LQ.
The 95\% CL mass limits for vector leptoquarks for $\kappa=1.0$ (gauge
coupling)
derived under this assumption are $M_{LQ}>240$ GeV/c$^2$ for
$\beta=1.0$; $M_{LQ}>240$ GeV/c$^2$ for $\beta=0.5$
(see Figure 2a).
The 95\% CL mass limits for vector leptoquarks for $\kappa=0$ (maximum
anomalous coupling) are $M_{LQ}>190$ GeV/c$^2$ for $\beta=1.0$; $M_{LQ}>185$
GeV/c$^2$ for  $\beta=0.5$ (see Figure 2b).

\widefigure{11pc}{\DZ\space 95\% confidence level lower limit on the first
generation vector leptoquark mass as a function of
$\beta$, for (a)$\kappa=1$ and (b)$\kappa=0$ (see text).
These limits assume the $same$ detection efficiency for the decay products
of scalar and vector leptoquarks.\label{pic2}}

\section{The CDF Detector}
    The CDF detector is described elsewhere [9]. The momenta of charged
particles are measured in the central tracking
chamber, which is surrounded by
a 1.4 T superconducting solenoidal magnet. This is surrounded by
electromagnetic
and hadronic calorimeters, which are used to identify jets. Outside the
calorimeters, drift chambers in the region $|\eta|<1.0$ provide muon
identification.

\section{Search for Second Generation Leptoquarks in CDF}

A second generation leptoquark $(S_2)$ would decay into a muon and a quark
with branching fraction $\beta$, or into an muon neutrino
and a quark, with branching fraction $(1-\beta)$. CDF has made a search
for $S_2\bar S_2$ pairs in the decay channel where both leptoquarks decay into
muon+quark. The experimental signature is 2 muons + 2 jets. The CDF
trigger sample, based on $19.3\pm 0.7 \ {\rm pb}^{-1}$ of data from the
1992-93 Tevatron run, was 7,958 events.
The offline requirements were:
\begin{itemize}
       \item [1)] two  good, isolated central muons with $p_t>20$ GeV/c
       \item [2)] two jets with $E_t>20$ GeV
\end{itemize}
\includegraphics{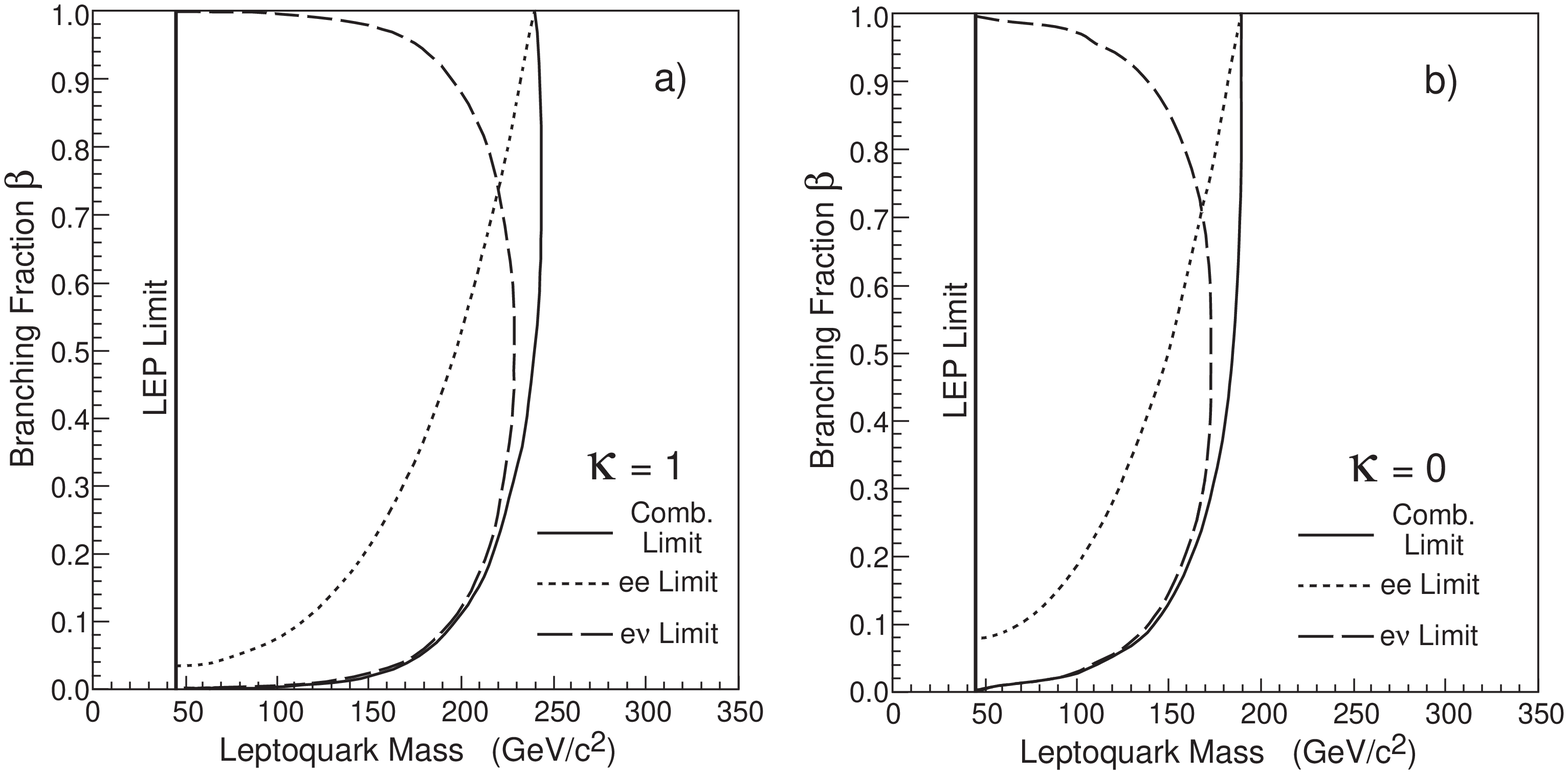}

After these cuts, 7 events remained. To remove background from $Z$ production,
events with $75<M(\mu\mu)<105$ GeV/c$^2$ were rejected. Two events remained
after this cut, with dimuon invariant masses of 18.9 GeV/c$^2$ and 57.9
GeV/c$^2$.
Background processes include Drell-Yan production of two
muons with two jets, $t\bar t$ production in the dimuon channel,
$Z\to \tau\bar {\tau}$,
 and fake muons. The total expected background is $1.35\pm 0.50$ events.

\section{Second Generation Scalar Leptoquark Mass Limits}
 Two candidate events are observed, consistent with the total background
of 1.35$\pm$0.50 events. The signal detection efficiencies were determined
using the ISAJET Monte Carlo program  [5] with CTEQ2L structure functions [10]
followed by a CDF detector simulation. The total efficiency ranges from
1.2\% for 40 GeV/c$^2$ leptoquarks to 16.8\% for 100 GeV/c$^2$, and is 12.5\%
for 120 GeV/c$^2$. Using these efficiencies and the observed number of events,
95\% CL limits on the cross-section times branching ratio were obtained.
By comparing these with the theoretical prediction for the cross-section
which is based  on ISAJET with CTEQ2L structure functions, lower limits
on the leptoquark mass as a function of $\beta$ were obtained. These are shown
in Figure 3. The 95\% CL lower limit on the mass of a second generation
scalar leptoquark for $\beta=1.0$ is 133 GeV/c$^2$, and the lower limit for
$\beta=0.5$
is 98 GeV/c$^2$.

\begin{figure}[htp]
\vspace*{ 11pc}
 \caption{\label{pic3}CDF 95\% confidence level lower limit on the
  second generation scalar leptoquark mass as a function of $\beta$.}
\end{figure}

\includegraphics{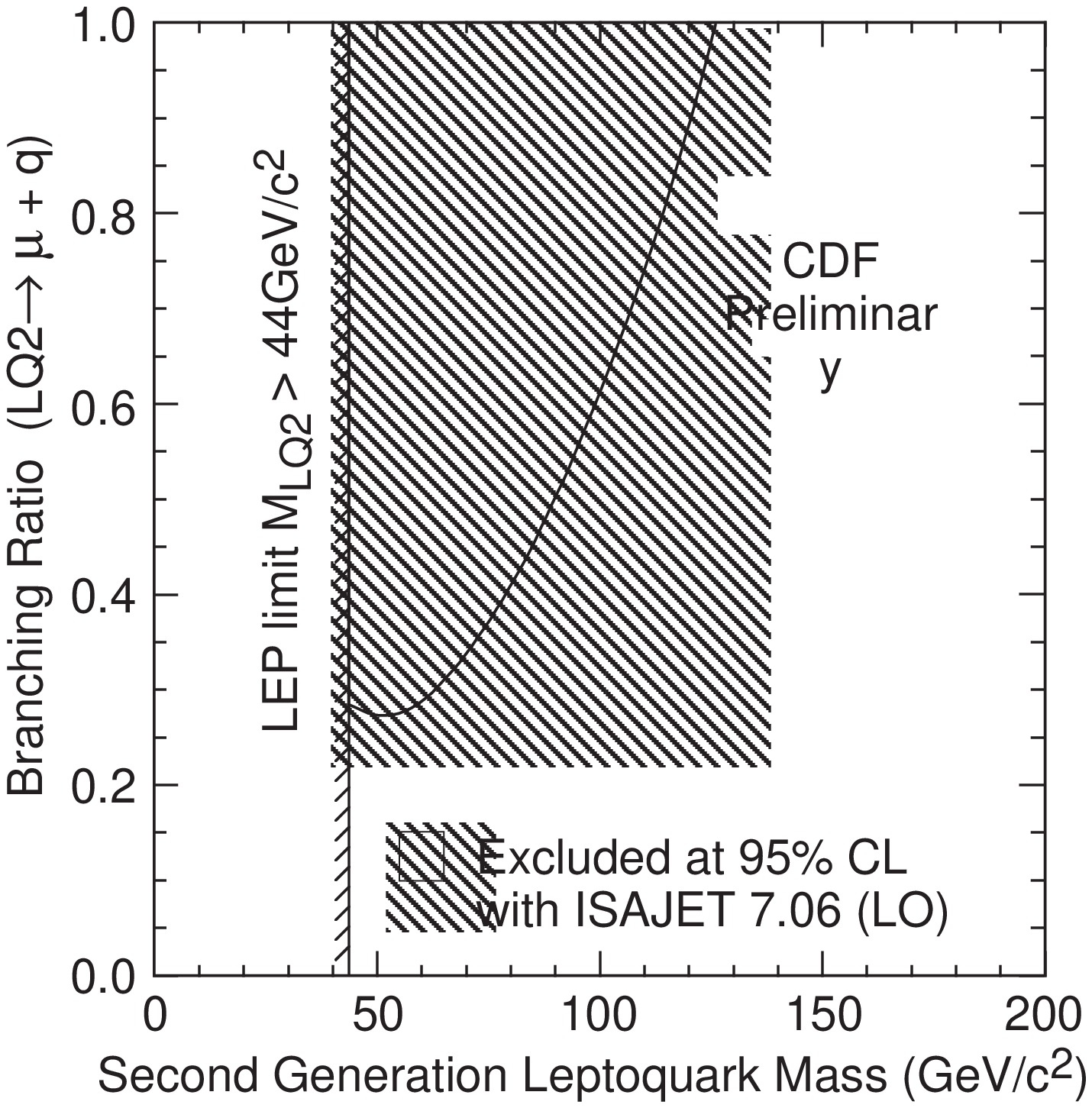}

\section{Minimal Supersymmetric Standard Model}
      One of the simplest supersymmetric extensions of the standard model (SM)
is the Minimal Supersymmetric Standard Model (MSSM) [11]. Supersymmetry
is a spacetime symmetry which relates bosons to fermions and introduces
supersymmetric partners (sparticles) for all the SM particles.
$R$-parity, the SUSY multiplicative quantum number, is defined as $R=+1$
for standard model particles and $R=-1$ for sparticles.
We assume that $R$-parity is conserved,
which implies that sparticles are produced in  pairs and decay to the stable
Lightest Supersymmetric Particle (LSP), which is usually assumed to be
the lightest Neutralino (\Zone).

\section{Search for Squarks and Gluinos}

  The experimental signature for squarks and gluinos is jets and/or leptons
and \etmiss, since the LSP does not
interact in the detector. In the \DZ\space search, 3 or more jets were
required.
Events with leptons were rejected to reduce the background from $W$ and $Z$
leptonic decays. The sample was 9,625 events from an integrated
luminosity of $13.4\pm 1.6 \ {\rm pb}^{-1}$. Offline requirements were:

\begin{itemize}
       \item [1)] a single interaction
       \item [2)] \etmiss \ $>75$ GeV
       \item [3)] three or more jets with $E_t>25$ GeV passing jet quality cuts
       \item [4)] reject electrons and muons
       \item [5)] no jet-\etmiss \ correlations
\end{itemize}
Of the 17 events surviving these cuts,
one event was rejected because it contained a muon consistent with a cosmic
ray, and two other events were rejected because their large \etmiss \ was
caused
by vertex reconstruction errors. The final candidate data sample contained
14 events, consistent with the
${18.5\pm1.9}^ {+7.6}_{-7.1}$ background events expected from
$W+2,3$ jets and QCD.

\section{MSSM Signal Simulation}

The MSSM model was used for the signal calculation, assuming SUGRA-inspired
degeneracy of squark masses [12]. Only squark and gluino production were
considered, no slepton or stop production.
 The mass of the top quark was set to 140 GeV/c$^2$.
To further specify the parameters,
the following values of MSSM parameters were used:
\begin{itemize}
       \item [1)] $\tan\beta=2.0$ (ratio of the Higgs vacuum expectation
values)
       \item [2)] $M(H^+)=500$ GeV/c$^2$ (mass of the charged Higgs)
       \item [3)] $\mu=-250$ GeV (Higgsino mass mixing parameter)
\end{itemize}

\begin{figure}[htp]
\vspace*{ 12pc}
 \caption{\label{pic4}\DZ, CDF, LEP and UA1/UA2 squark and gluino mass
limits as a function of squark and gluino mass.}
\end{figure}


\includegraphics{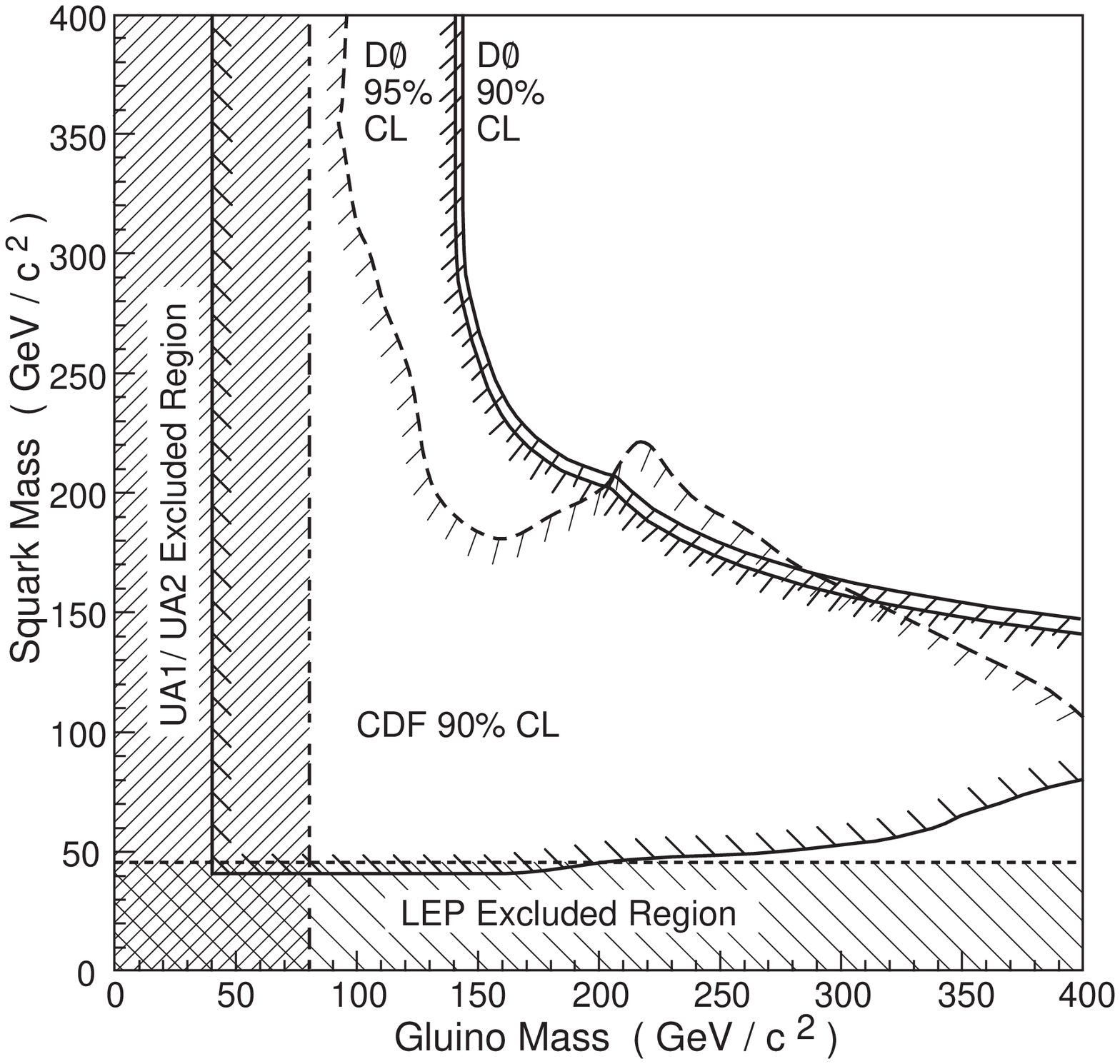}

Squark and gluino events were simulated using the ISASUSY event generator [13]
and processed through the \DZ\space triggering,
detector simulation, and reconstruction programs.
The detection efficiency, $\epsilon$, was determined for 28 squark and gluino
mass combinations using masses between 100 and 400 GeV/c$^2$.
As an example, for 200 GeV/c$^2$ equal mass squarks and gluinos, $\epsilon
= 18.6\%$.
For $m_{\tilde g}$ = 150 GeV/c$^2$ and
$m_{\tilde q}$ = 400, $\epsilon = 6.3 \%$.
And for $m_{\tilde g}$ = 400 GeV/c$^2$ and
$m_{\tilde q}$ = 150, $\epsilon = 8.3 \%$.
A combination of linear fitting and linear interpolation was used to
find efficiencies between grid points.

\section{Calculation of Mass Limits from Cross Sections}
Using these signal detection
efficiencies, the luminosity and the number of visible events above SM
background, a 95\% CL upper limit cross section
was determined. This was compared with
a leading order theoretical cross section [13] to determine the
lower mass limit for each squark and gluino mass combination.
For heavy squarks, a lower gluino mass limit
of 146 GeV/c$^2$ was obtained, and for equal squark and gluino masses,
a mass limit of 205 GeV/c$^2$ was obtained at the 95\% CL.

\section{Acknowledgements}

I am grateful to
the members of the CDF and \DZ\space collaborations for their hard work.
Special thanks are due to D. Norman, \DZ,  for the leptoquark
analysis, M. Paterno,
\DZ,\space for the squark and gluino analysis, and to S. Park, CDF, for the
second generation leptoquark analysis.

\end{document}